\documentclass[pre,aps,twocolumn,superscriptaddress,amsmath,floatfix]{revtex4}
\usepackage{graphicx}
\begin{document}
\title{Fluctuation induced attraction between adhesion sites of supported membranes}
\author{Oded Farago}
\affiliation{Department of Biomedical Engineering, Ben Gurion University,
Be'er Sheva 84105, Israel}
\begin{abstract}
We use scaling arguments and coarse grained Monte Carlo simulations 
to study the fluctuation mediated interactions between a pair of
adhesion sites of a bilayer membrane and a supporting surface. We find
that the potential of mean force is an infinitely long range attractive
potential that grows logarithmically with the pair distance $r$:
$\phi(\vec{r})/k_BT=c\ln r$, where the constant $c=2$ and 
$c=1$ for non-stressed and stressed membranes, respectively. When, 
in addition to excluded volume repulsion, the membrane
also interacts with the underlying surface through a height-dependent 
attractive potential, the potential $\phi(\vec{r})$ is screened at
large pair distances.
\end{abstract}
\maketitle

Supported lipid membranes are useful and important model systems for
studying cell membrane properties and membrane mediated processes
\cite{sackmann:1996,girard:2007}. Placing a membrane on a flat
substrate allows for the application of several different surface
sensitive techniques, including atomic force microscopy, x-ray and
neutron diffraction, ellipsometry, nuclear magnetic resonance, and
others \cite{rinia:2006}.  With the aid of biochemical tools and
generic engineering, supported membranes can be functionalized with
various membrane-associated proteins \cite{tanaka:2006}. One
attractive application of supported membranes is the design of phantom
cells exhibiting well defined adhesive properties and receptor
densities \cite{salafsky:1996}. Using advanced imaging techniques,
detailed information can be obtained about the structure of the
adhesion zone between the receptor-functionalized supported membrane
and ligand-containing vesicles that can bind to the supported membrane
\cite{kloboucek:1999,kaizuka:2004}. These studies provide insight into
the specific (ligand-receptor) and nonspecific interactions during
cell adhesion \cite{smith:2006,ananth:2007}. Understanding these
interactions is crucial for the development of drug delivery systems
that depend on efficient adhesion between a liposome and the plasma
membrane of the target cell.

Adhesion is an immensely complex process involving many
physio-chemical and biomolecular factors \cite{cerca:2005}. Many
aspects of this process, ranging from the cooperativity in adhesion
cluster formation to the influence of stochastic processes such as the
ligand-receptor reaction kinetics, have been and continue to be
studied theoretically using various models (see, e.g., recent reviews
in refs.~\cite{smith:2007,weikl:2009}). In light of this extensive
theoretical effort, it is surprising that there is still no
satisfactory answer to one of the most fundamental problems associated
with adhesion, namely the characterization of the membrane-mediated
interaction between adhesion sites. Detailed knowledge of the strength
and range of these interactions is essential for a better
understanding of the role they play during the self-assembly of
adhesion zones. The lack of theoretical studies of membrane mediated
interactions between adhesion sites is in striking contrast with the
extensive literature existing on membrane-mediated interactions
between transmembrane proteins. In the latter case, the origin of the
interactions is the ability of two proteins to position themselves in
a manner which minimizes the total bending elastic energy of the
deformed membrane \cite{bruinsma:1996}. In addition, the influence of
the proteins on the membrane thermal fluctuations leads to
membrane-mediated interactions between them which are analogous to
Casimir forces between conducting plates \cite{kardar:1999}. These
interactions fall off with the protein pair separation as $1/r^4$
\cite{bruinsma:1996,kardar:1999,park:1996} and, therefore, at large
distances they are considerably larger than Van der Waals and screened
electrostatic interactions which decay much faster with $r$. Since
adhesion sites between membranes or between a membrane and a surface
represent a different type of ``constraint'' on the shape of the
membrane, one can expect Casimir-like interactions to exist between
them as well. Below, we explore these interactions for a pair of
adhesion sites between a membrane and a flat, impenetrable, surface
and show that these interactions are of effective infinite range.

Consider a membrane of linear size $L$ with bending rigidify $\kappa$
and surface tension $\sigma$, which may also experience a
height-dependent harmonic confining potential (whose second derivative
with respect to the height is $\gamma$) due to the presence of a nearby
flat surface. Let $h(\vec{r})$ be the height function of the
membrane, which vanishes at the points where the membrane is attached
to the surface and takes positive vales everywhere else. The total
energy of the membrane is, thus, is given by the effective Hamiltonian
\begin{equation}
{\cal H} = \int\,
\frac{1}{2}\left[\kappa\left(\nabla^2 h \right)^2+\sigma\left(\vec{\nabla} h 
\right)^2+\gamma h^2\right]\Phi\left(h\right)
d^2\vec{r},
\label{eq:helfrich}
\end{equation}
where $\Phi$ represents the hard wall constraint ($\Phi=1$ for $h\geq
0$, and $\Phi=+\infty$ for $h<0$), and the integration is taken over
the cross sectional (projected) area of the membranes $A_p\sim
L^2$. Let us first consider the case where $\sigma=0$ and $\gamma=0$
in Eq.~(\ref{eq:helfrich}). In a previous publication we studied the
behavior of a membrane with one attachment site to the surface
\cite{farago:2008a}. We found, quite unexpectedly, that the attachment
of the membrane to a flat surface at only single adhesion point does
not modify the spectrum of thermal fluctuation of the membrane. The
only effect of the attachment is to eliminate the membrane
translational degree of freedom by enforcing that the global minimum
of the height function $h(\vec{r})$ is achieved at the point of
contact with the surface. Without the surface (i.e., for a freely
fluctuating membrane), the manifold could be translated horizontally
and the global minimum could be transferred to any place within the
cross sectional area $\vec{r}\in A_p$. The attachment free energy cost
is, therefore, $\Delta F=k_BT\ln(A_p/l^2)=2k_BT\ln(L/l)$, where $k_B$
is Boltzmann's constant, $T$ is the temperature, and $l$ is some
microscopic length scale of the order of the bilayer thickness. The
scaling behavior of $\Delta F$ with $L$ can be also obtained by
noting that because the single attachment point leaves the spectrum of
thermal fluctuation unchanged, the mean height of the membrane above
the surface increases as $u(r)=\langle h(r)\rangle\sim
r(\sqrt{k_BT/\kappa})$ with $r$, the distance from the pinning site
\cite{farago:2008a,bruinsma:1994}. Helfrich showed that, as a result
of the collisions between the membrane and the surface, there is an
effective interaction energy per unit area: $V(r)\sim (k_BT)^2/\kappa
u(r)^2\sim (k_BT)/r^2$ \cite{helfrich:1978}. By integrating this
energy density over the projected area of the membrane, one finds that
\begin{equation} 
\Delta F = \int\, V(r)d^2\vec{r}
=Ck_BT \ln\left(\frac{L}{l}\right).
\label{eq:attachenergy}
\end{equation}
An equation very similar to Eq.~(\ref{eq:attachenergy}) has been
previously derived in ref.~\cite{bruinsma:1994} in the context of
self-assembly of membrane junctions. In that reference, the mean field
free energy per adhesion site was found to have a logarithmic
dependence on the mean distance between sites. This result has been
interpreted as a renormalization of the temperature downward.  Our
finding that $C=2$, implies that the attachment free
energy (\ref{eq:attachenergy}) exactly cancels the mixing entropy term
of the adhesion sites and, therefore, the renormalized temperature
$T_r=0$. The inability of this simple mean field calculation to
predict whether the adhesion sites tend to aggregate ($T_r<0$) of
segregate ($T_r>0$), emphasizes the need for a more detailed analysis
of the membrane mediated interactions that also takes into account
their many-body nature. We leave most of the discussion of many-body
effects to a future publication and focus here on the pair correlation
function between adhesion sites.

The attachment free energy (\ref{eq:attachenergy}) is distributed
within a volume ${\cal V}\sim L^2\Delta _0$, where $\Delta_0=u(L)\sim
L\sqrt{k_BT/\kappa}$~is the mean height of the membrane. It,
therefore, seems reasonable to speculate that disjoining pressure
between the membrane and the surface scales as
\begin{equation}
P\sim \Delta F/{\cal V}\sim \sqrt{k_BT\kappa}/L^3\ln(L/l).
\label{eq:pressure1}
\end{equation} 
The pressure between the membrane and the underlying surface is not
uniform, however, but rather decreases with $r$ because points on the
manifold that are closer to the attachment site tend to collide more
frequently with the surface. Defining the distance-dependent pressure,
$P(r)$, the mean disjoining pressure can be calculated by: $P\sim
(1/L^2)\int_l^L rP(r) dr$, which can be reconciled with
Eq.~(\ref{eq:pressure1}) by assuming that $P(r)\sim
\sqrt{k_BT\kappa}/Lr^2$. Since the pressure is caused by collisions
between the membrane and the surface, one may conclude that the
probability density that the membrane comes into contact with the
surface at distance $r$ from the attachment point has the same scaling
behavior as $P(r)$:
\begin{equation}
\Pi\left[h\left(\vec{r}\right)=0\right]\sim P\left(\vec{r}\right)\sim 
\frac{1}{r^2}.
\label{eq:probability}
\end{equation}

Let us now turn to the problem of a supported membrane with two
adhesion points.  Let $\vec{r}=\vec{r}_0$ denote the position of the
second adhesion point within the cross sectional area of the membrane,
while the first adhesion point is fixed at the origin,
$\vec{r}=\vec{0}$. The potential of mean force between the adhesion
sites is defined as $\phi(\vec{r}_0)=-k_BT\ln(g(\vec{r}_0))$, where
$g(\vec{r}_0)$ is the pair distribution function expressed as a
function of the coordinate of the second adhesion site. By definition,
$g(\vec{r}_0)=Z(\vec{0},\vec{r}_0)/Z(\vec{0})$, where $Z(\vec{0})$ and
$Z(\vec{0},\vec{r}_0)$ denote the partition functions of membranes
with one (at $\vec{0})$ and two (at $\vec{0}$ and $\vec{r}_0$)
adhesion sites, respectively. However, the ratio
$Z(\vec{0},\vec{r}_0)/Z(\vec{0})$ is also equal to
$\Pi\left[h\left(\vec{r}_0\right)=0\right]$, the probability density
that a configuration with an adhesion site at $\vec{r}=\vec{0}$ makes
contact with the surface at $\vec{r}_0$ as well.
We thus conclude that
$g(\vec{r}_0)=\Pi\left[h\left(\vec{r}_0\right)=0\right]$, and together
with Eq.~(\ref{eq:probability}), we arrive to the following scaling
result for the pair correlation function
\begin{equation}
g(\vec{r}_0)\sim 1/r_0^2.
\label{eq:pairfunc}
\end{equation}
From this we find that the potential of mean force between the two
adhesion sites is an infinitely long range attractive potential that
grows logarithmically with the pair distance $r_0$:
$\phi(\vec{r}_0)=-k_BT\ln(g(\vec{r}_0))=2k_BT\ln\left(r_0\right)$.

We tested the validity of Eq.~(\ref{eq:pairfunc}) by using constant
surface tension (frame tension) Monte Carlo (MC) simulations of a
coarse-grained implicit solvent bilayer model. The details of the
model (which is suitable for simulations of bilayer membranes at large
spatial and temporal scales) and simulations can be found in
ref.~\cite{farago:2008a}, where we discuss the problem of a membrane
with one attachment point to the surface. In the present work, we have
a flat surface and 2000 coarse grained lipids that form a square
bilayer patch of linear size $L$. Each lipid is represented by a short
string consisting of one head bead and two tail beads. The lipids
reside on one side of the surface. Two lipids are attached to surface
at their head beads. The location of one of these head beads is fixed
at the origin, while the second head bead is allowed to diffuse on the
flat surface. By sampling the position of the latter bead, the pair
distribution function can be computed and compared with the power law
distribution, Eq.~(\ref{eq:pairfunc}). There is, however, a problem
with this seemingly straightforward strategy. The simulation time must
be much longer than both (i) the typical relaxation time of the
longest bending mode and (ii) the typical diffusion time across the
membrane of the mobile adhesion point. Unfortunately, both these
characteristic times grow very rapidly with $L$, in a way which makes
the application of the standard Metropolis MC algorithm
impractical. To overcome this problem we used two ``tricks'': The
relaxation times of the thermal bending modes were reduced by applying
the recently proposed ``Mode Excitation MC'' (MEMC) scheme
\cite{farago:2008b}. The MEMC scheme utilizes collective update moves
that lead to fast excitation and relaxation of the long wavelength
modes. The problem arising from the slow diffusion of the mobile
adhesion point was solved by identifying the unpinned lipid whose head
group is located closest to the surface and introducing a new MC move
that attempts to place this lipid on the surface while lifting the
mobile pinned lipid away from the surface. More precisely, if the
height of the closest unpinned headgroup to the surface is $h_0>0$,
the move attempt consists of a simultaneous change in the heights of
all the beads of both the unpinned and pinned lipids - the former are
reduced by $h_0$, while the latter are increased by $h_0$. One can
easily verify that for this new type of MC move, detailed balance is
satisfied by the conventional Metropolis acceptance rule: $p({\rm
old}\rightarrow{\rm new})={\rm min}(1,\exp(-\Delta E/k_BT))$, where
$\Delta E$ is the energy change caused by the move attempt. The new
move allows the mobile pinning site to ``jump'' from from one place on
the membrane to another and enables efficient sampling of the pair
correlation function $g(\vec{r}_0)$ within a reasonable simulation
time.

\begin{figure}[t]
\begin{center}
\scalebox{0.325}{\centering \includegraphics{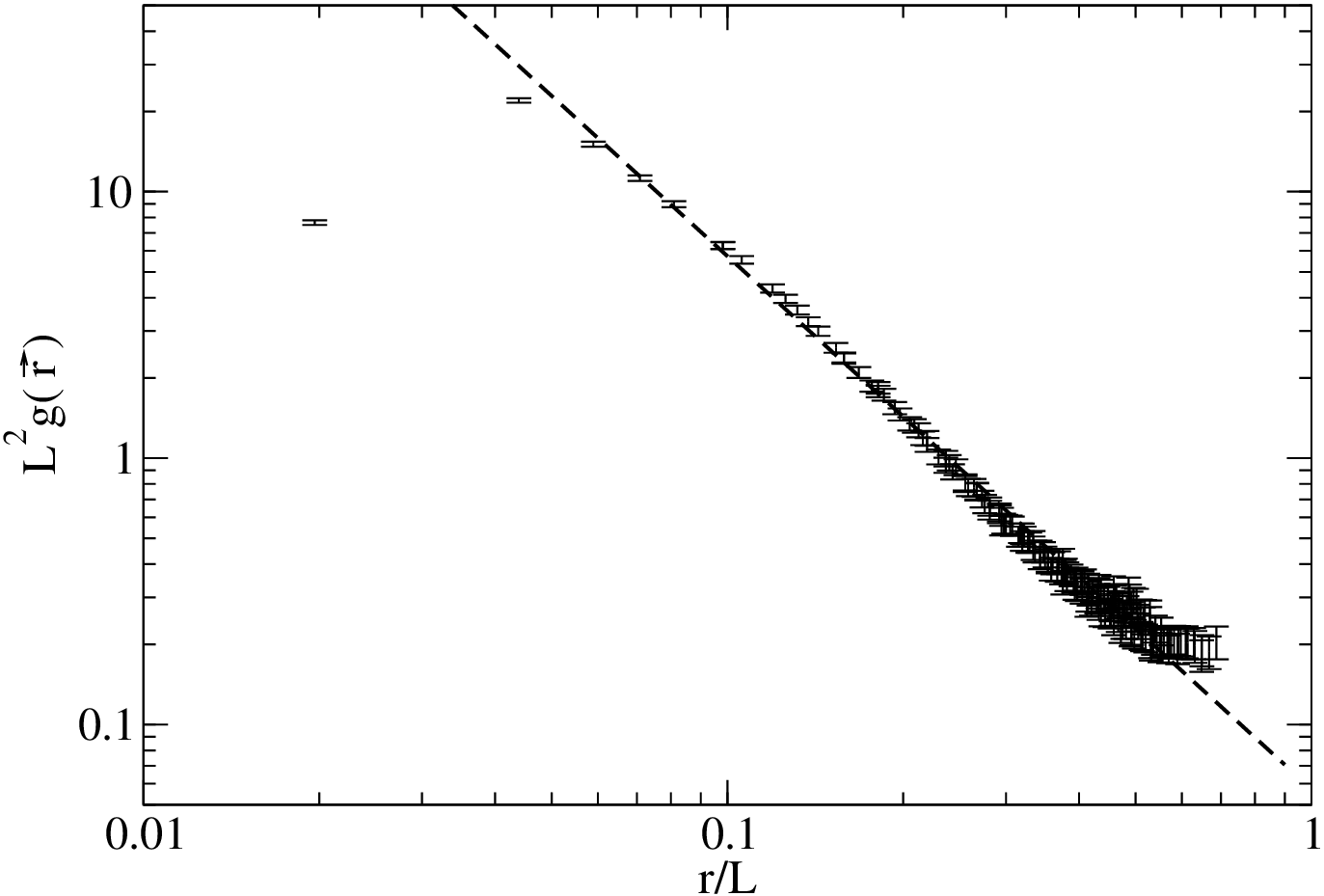}}
\end{center}
\vspace{-0.75cm}
\caption{The pair correlation function, $g\left(\vec{r}\right)$, of a
non-stressed membrane vs.~the pair distance $r$. The slope of the
dashed straight line is $-2$.}
\label{Figure1}
\end{figure}

Our results, which are shown in Fig.~\ref{Figure1}, agree very well
with the functional form predicted in Eq.~(\ref{eq:pairfunc}). The
slope of the straight line on the log-log plot is equal to
$-2$. Deviations from the power law behavior $g(\vec{r})\sim 1/r^2$
can be observed at small ($r/L<0.05$) and large ($r/L>0.5$) distances
only. At small separations, the molecular nature of the lipids becomes
important and the radial pair distribution function is dominated by
the depletion shells around the lipids. At large distances, the mobile
adhesion site reaches the center of the square membrane where it
becomes attracted not only by the fixed adhesion site but also by its
periodic images. In this regime, the many-bony nature of the
membrane-mediated potential must be taken into account. 

\begin{figure}[t]
\begin{center}
\scalebox{0.325}{\centering \includegraphics{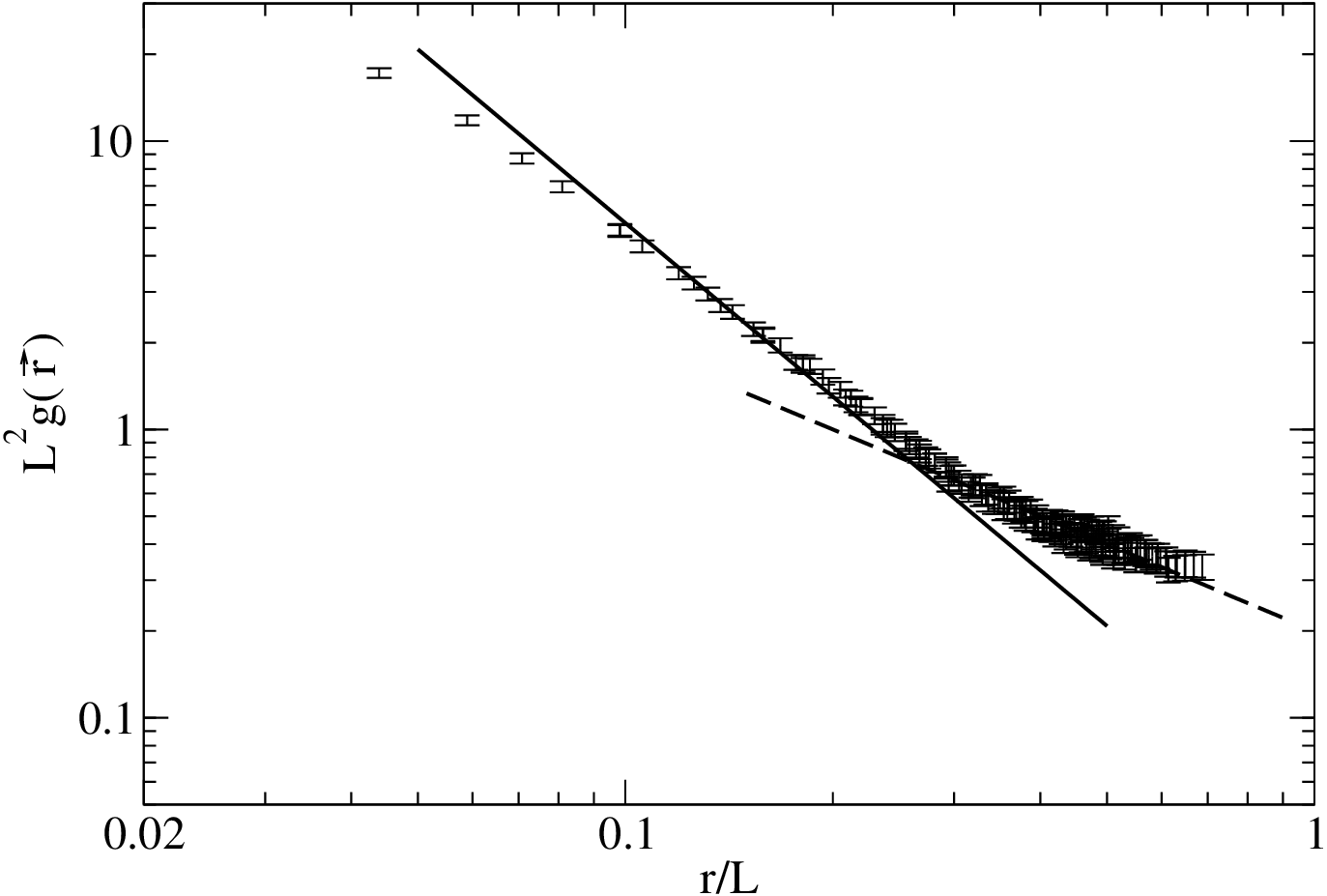}}
\end{center}
\vspace{-0.75cm}
\caption{The pair correlation function, $g\left(\vec{r}\right)$, of a
stressed membrane vs.~the pair distance $r$. The slopes of the solid
and dashed straight lines are $-2$ and $-1$, respectively.}
\label{Figure2}
\end{figure}

Let us now consider the case of a membrane which is also subjected to
lateral surface tension $\sigma>0$ (see Eq.~(\ref{eq:helfrich})). For
a stressed membrane, one can define the ``surface tension crossover
length'', $\xi_{\sigma}\sim \sqrt{\kappa/\sigma}$, which marks a
crossover between two regimes. On length scales much smaller than
$\xi_{\sigma}$, the thermal fluctuations are governed by the bending
elasticity of the membrane while the surface tension term in
Eq.~(\ref{eq:helfrich}) is negligible. Therefore, for
$r\ll\xi_{\sigma}$, one can expect Eq.~(\ref{eq:pairfunc}) for the
pair correlation function to hold. On length scales much larger than
$\xi_{\sigma}$, the surface tension term in Eq.~(\ref{eq:helfrich})
becomes dominant, which leads to the suppression of the long
wavelength thermal fluctuations. Consequently, for $r\gg\xi_{\sigma}$,
the decay of the pair correlation function should be slower than
predicted by Eq.~(\ref{eq:pairfunc}). The behavior of $g(\vec{r})$ in
this regime can be derived using the following argument: Let
$h(\vec{r})$ be the height function of the stressed membrane. Define
the function $H(\vec{r})$ such that $\kappa\left(\nabla^2 H
\right)^2=\sigma(\vec{\nabla} h)^2$. The manifold depicted by the
function $H$ represents a non-stressed bilayer membrane, and from
Eq.~(\ref{eq:probability}) we have that
$\Pi\left[H\left(\vec{r}\right)=0\right]\sim 1/r^2$. At large $r$, the
two functions can be related by the simple scaling relation $\kappa
H^2/r^4 \sim \sigma h^2/r^2$, i.e., $h(r)\sim (\xi_{\sigma}/r)H(r)$
\cite{remark}. This scaling relationship implies that the probability
density $\Pi\left[h\left(\vec{r}\right)=0\right]\sim
\left(r/\xi_{\sigma}\right)\Pi\left[H\left(\vec{r}\right)=0\right]$. As
previously noted, the function
$\Pi\left[h\left(\vec{r}\right)=0\right]$ is, in fact, the pair
distribution function, which leads us to conclude that in the
surface-tension dominated regime, $r\gg\xi_{\sigma}$: $g(\vec{r})\sim
(r/\xi_{\sigma})1/r^2\sim 1/r$. The potential of mean force,
$\phi(\vec{r})=-k_BT\ln(g(\vec{r}))=k_BT\ln\left(r\right)$, simply
decreases to half of the value of $\phi(\vec{r})$ in non-stressed
membranes. Notice that this form is independent of $\sigma$ whose
magnitude influences only the crossover length $\xi_{\sigma}$ between
the two scaling regimes. Fig.~\ref{Figure2} shows our simulation
results for a membrane with surface tension $\sigma=3.6 k_BT/l^2$,
where the microscopic length scale $l$ is taken as the length of the
three-bead model lipid ($l\sim 2\ {\rm nm}$. The linear size of the
membrane patch in our simulations $L\sim 12.5l$). The two scaling
regimes with $g(\vec{r})\sim 1/r^2$ for small $r$ and $g(\vec{r})\sim
1/r$ for large $r$ can be clearly seen. In a previous paper we
measured the bending rigidity of the membrane and found $\kappa\sim
8k_BT$ \cite{farago:2008b}. This gives $\xi_{\sigma}\sim 1.5 l\sim
0.12 L$, which is consistent with Fig.~\ref{Figure2} where the
crossover between the scaling regimes takes place around $r\sim
0.25L$.

\begin{figure}[t]
\begin{center}
\scalebox{0.325}{\centering \includegraphics{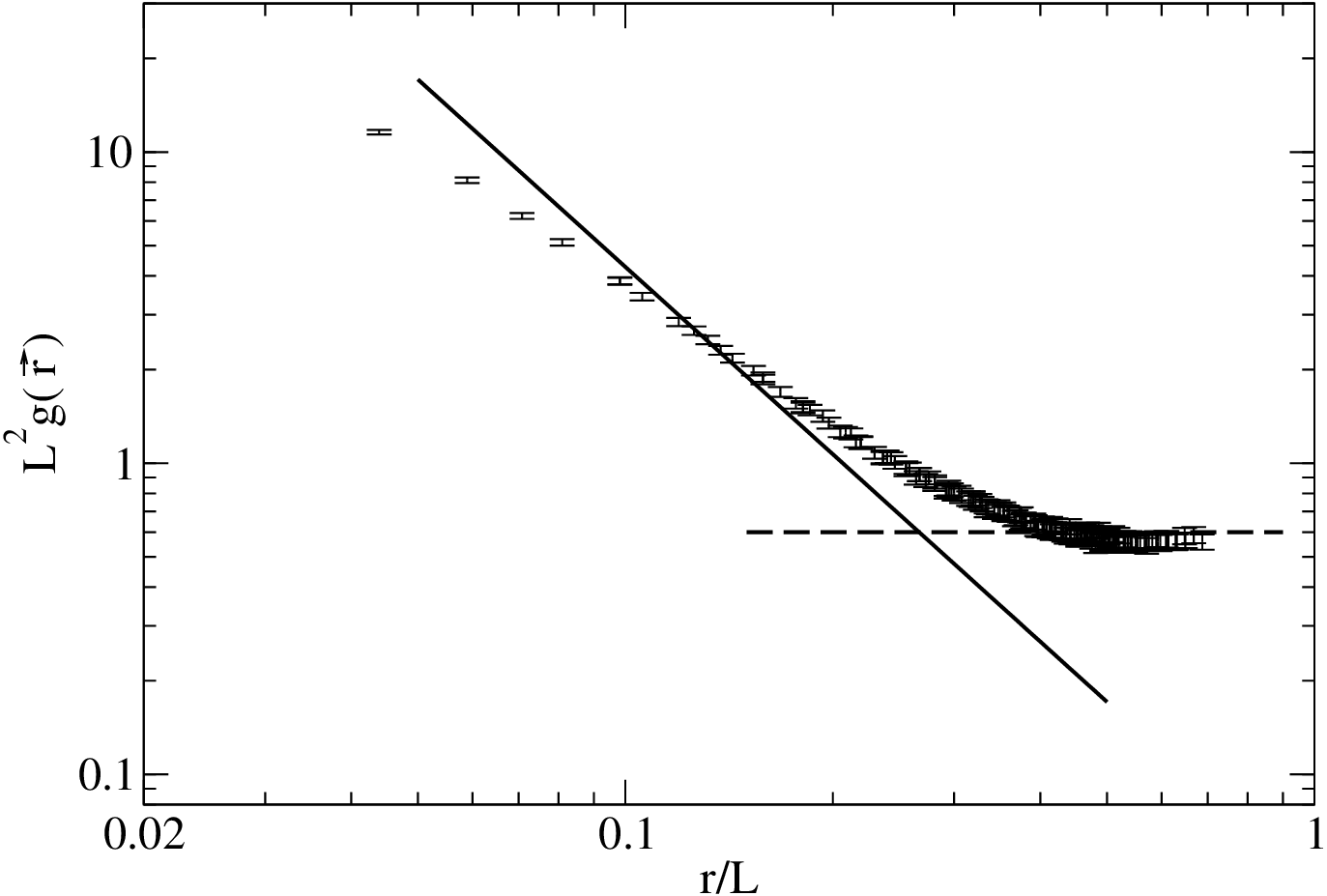}}
\end{center}
\vspace{-0.75cm}
\caption{The pair correlation function, $g\left(\vec{r}\right)$, of a
membrane experiencing harmonic confining potential vs.~the pair
distance $r$. The slopes of the solid and dashed straight lines are
$-2$ and $0$, respectively.}
\label{Figure3}
\end{figure}

For a non-stressed membrane experiencing a harmonic confining
potential ($\gamma>0$ in Eq.~(\ref{eq:helfrich})), the ``harmonic
confinement crossover length'', $\xi_{\gamma}\sim
(\kappa/\gamma)^{1/4}$, can be defined which marks the transition
between two scaling regimes. For $r\ll\xi_{\gamma}$, the thermal
fluctuations are dominated by the bending rigidity term in the
Hamiltonian and, therefore, $g(\vec{r})\sim 1/r^2$.  For
$r\gg\xi_{\gamma}$, the fluctuation spectrum is dominates by the
harmonic confinement term. Since this term is a local one, the
influence of the adhesion site is screened and pair correlation
function saturates to a constant value, $g(\vec{r})\sim r^0$, which
means that fluctuation mediated force between the adhesion sites
vanishes \cite{remark2}. To verify these predictions, we added an
energy term for each lipids in our model, $E=(1/2)\gamma^*h_l^2$,
which is proportional to the square of the height $h_l$ of the
lipid. The variable $\gamma^*$ is related to $\gamma$ in
Eq.~(\ref{eq:helfrich}) by $2\gamma^*/a=\gamma$, where $a$ is the
cross-sectional area per lipid and the factor 2 is due to the two
leaflets of the membranes. Here, we take $\gamma^*=0.072$, which
together with the previously computed value $a\simeq 0.15l^2$
\cite{farago:2008b}, gives $\gamma\simeq 0.096\ k_BT/l^4$.  This
yields, $\xi_{\gamma}\sim 3l\sim 0.25L$. The results, presented in
Fig.~\ref{Figure3}, show the two scaling regimes for small and large
$r$, where the crossover between them occurs at $r\sim 0.3L$.

To conclude, we have calculated the membrane-mediated interactions
between two adhesion site of a bilayer membrane and a supporting flat
surface.  We found that the potential of mean force is an infinitely
long range attractive potential that grows logarithmically with the
pair distance $r$: $\phi(\vec{r})/k_BT=c\ln r$. The constant $c$ takes
three possible values depending on which term in
Eq.~(\ref{eq:helfrich}) dominates the effective Hamiltonian of the
system: $c=2$ in the bending-rigidity dominated regime (which always
prevails at small pair separations), $c=1$ in the surface-tension
dominated regime, and $c=0$ in the regime where the dominant term is
of the harmonic confinement. It is important to note that even when
$c=2$ at all pair separations (i.e., for $\sigma=0$ and $\gamma=0$ in
Eq.~(\ref{eq:helfrich})), the membrane-mediated attractive potential
is not strong enough to bind the pair of adhesion sites. The pair
correlation function in this case, $g(\vec{r})\sim 1/r^2$, and the
mean pair separation increases with the size of the system: $\langle
r\rangle =\int_l^L r^2g(r)dr/\int_l^L rg(r)dr\sim L/\ln L$. This, however,
does not render the fluctuation mediated interactions between adhesion
sites unimportant. These interactions may very well provide a powerful
aggregation mechanism of receptor-ligand binding domains. In order to
correctly analyze the aggregation behavior of an ensemble of adhesion
sites, one must take into account the many-body nature of the
fluctuation induced interactions between them. Such a study is
currently underway.



\begin{thebibliography}{99}
\bibitem{remark2} This argument also holds for potentials of the form
$V=(1/2)\gamma(h-h^*)^2$ as well. In this case, the screening takes
place on length scales much larger than (twice) the ``healing length''
of each adhesion point, which is the typical length from the adhesion
site at which the membrane returns to the equilibrium spacing,
$h=h^*$, from the surface.
\end{thebibliography}
\end{document}